\documentclass[12pt,preprint]{aastex}

\usepackage{natbib}
\slugcomment{Accepted for publication in ApJ}
\shorttitle{The value of the fine structure constant over cosmological times}
\shortauthors{Guti\'errez \& L\'opez-Corredoira}

\begin{document}
\title{The value of the fine structure constant over cosmological times}

\author{C. M. Guti\'errez and M. L\'opez-Corredoira}
\affil{Instituto de Astrof\'\i sica de Canarias, 38205 La Laguna, Tenerife, SPAIN}

\email{cgc@iac.es}
\begin{abstract}

The optical spectra of objects classified as QSOs in the SDSS DR6 are analyzed with the
aim of determining the value of the fine structure constant in the past  and then check
for possible changes in the  constant over cosmological timescales. The analysis is done
by measuring the position of the fine structure lines of the [OIII] doublet ($\lambda
\lambda 4959$ and $\lambda\lambda 5008$) in QSO nebular emission. From the sample of 
QSOs at redshifts  $z<0.8$ a sub sample was selected on the basis of the amplitude and
width of the [OIII] lines. Two different method were used to determine the position of
the lines of the [OIII] doublet, both giving similar results. Using a clean sample
containing 1568 of such spectra, a value of  $\Delta \alpha /\alpha=(+2.4\pm 2.5)\times
10^{-5}$ (in the range of redshifts $z\sim 0-0.8$) was determined. The use of a larger
number of spectra allows  a factor
$\sim 5$ improvement on previous constraints based on the same method. On the whole, we
find no evidence of changes in $\alpha$ on such cosmological timescales. The mean
variation compatible with our results is $1/ <t>\Delta \alpha/\alpha=(+0.7\pm 0.7)\times
10^{-14}$ yr$^{-1}$. The analysis was extended to the [NeIII] and [SII]
doublets, although their usefulness is limited due to the fact that all these doublets
in QSOs tend to be fainter than [OIII], and that some of them are affected by
systematics.
\end{abstract}

\keywords{atomic data--line identification--line profiles--quasars: emission lines}

\section{Introduction}

In Kaluza-Klein and super string theories that try to unify all the fundamental interactions, the
constants of nature are functions of a low mass dynamical scalar field, which slowly  changes
over cosmological timescales \citep{uza03}. Astrophysics offers a possible test to
constrain the parameters of such theories by directly measuring  the values of such constants
through the comparison of properties of objects at different evolutionary epochs of the Universe. Reviews of
the techniques used have been presented by \citet{lan08}, \citet{kan08} 
and others. The biggest observational effort to constrain the values of  physical constants
in the past has been directed at the fine structure constant ($\alpha \equiv e^2/\hbar c$), which
plays a fundamental role in the characterization of electromagnetic interaction.  Several methods
have been developed  based on the analysis of cosmic microwave background data \citep{nak08}, Big
Bang nucleosynthesis \citep{ich02}, and the fine structure splitting of several
atomic lines in QSO spectra (see below). A review on the techniques and observational 
constraints can be found in \citet{gar08}. 

Here, we  mention only those methods based on  fine structure splitting. The splitting
ratio ($\lambda_2-\lambda_1)/(\lambda_2+\lambda_1)$  at two different epochs  gives the
relative difference in $\alpha$ between these two epochs. It is shown  \citep{uza03} that

$$ \frac{\Delta \alpha}{\alpha}(z)=\frac{1}{2}\left
\{\frac{[(\lambda_2-\lambda_1)/(\lambda_2+\lambda_1)]_z}
{[(\lambda_2-\lambda_1)/(\lambda_2+\lambda_1)]_0}-1\right \} $$ where $\lambda_2$ and $\lambda
_1$ are the wavelengths of the pairs of the doublet, and the subscripts  $z$ and $0$ refer to the
values at redshift $z$ and local respectively.

The method has been employed to measure the relative separation of absorption lines
in the spectra of QSOs, including the alkali doublet  \citep{bah67}, and the many
multiplet method  (MML), \citep{web99}. The best constraints obtained using the 
alkali doublet method are  those by \citet{cha05} [$\Delta \alpha/\alpha=(0.15\pm
0.44)\times 10^{-5}$] over the range $1.59\le z \le 2.32$]. The MML simultaneously
analyzes many doublets of many atomic species, and  due to the large number of lines
used, higher  precision than with the alkali doublet method. 

The MML is  the only
method that has resulted in claims for the detection of variation of $\alpha$
\citep{web99,mur03} $\Delta \alpha/\alpha=(-0.574\pm 0.102)\times 10^{-5}$ in the
range $0.2\le z\le 3.7$, although these results are controversial \citep{cha06}, (but
see also \citep{mur08}  on the opposite view), and are at odds with similar studies
\citep{cha04,sri04,lev07}. 

The MML method  suffers from several uncertainties and
might be affected by possible systematics, as was pointed out by \citet{bah04} among
others.  Another group \citep{lev05}  has developed a slight modification of the MML
method in which  only one atomic ion (Fe II) is used, avoiding many of the
assumptions and uncertainties inherent to the MML method. Their latest analysis 
\citep{lev07,mol08}  determined $\Delta \alpha/\alpha=(-0.12\pm 1.79)\times 10^{-6}$
at $z=1.15$ and ${\Delta \alpha /\alpha }$ = $(5.4\pm 2.5)\times 10^{-6}$  at
$z=1.84$ although the authors cautiously stated that there could be some unnoticed
systematic effects that might challenge that interpretation.

In this paper, we use the [OIII] nebular emission lines in QSOs to constrain  past variations in
$\alpha$. The method was proposed by \citet{bah65}, and was later
applied by \citet{bah04} and  \citet{gru05}. Because the analysis is
based  on a pair of lines only, the constraints on $\Delta \alpha/\alpha$ are not as strong as those
 obtained with the other methods mentioned above, but has the advantage that it is more
transparent and less subject to systematics. The method is based on the same element and
level of ionization, and both lines  originate in the same upper energy level, so  the analysis
is  quite independent of the physical conditions of the gas where the [OIII] lines originate.
It therefore represents a good alternative
for determining the value of $\alpha$ on a firm basis. \citet{bah04} analyzed  the QSOs of
the Early Data Release of SDSS and built a clean sample of 42 QSOs in the range $0.16\le z \le
0.80$, from which they derived $\Delta \alpha/\alpha=(+0.7\pm 1.4)\times 10^{-4}$. Here, we use the
QSOs in the latest release of  the SDSS to improve such constraints significantly.

\section{Sample selection and methodology}

The \emph {SDSS-DR6 Catalog Archive
Server}\footnote{http://www.sdss.org/dr6/access/index.html\#CAS} contains 77\,082 objects
classified as QSOs. By SQL queries we  downloaded the extracted 1d spectra of all of
them. These spectra have a resolution $\sim 2000$ and the range covered is 
$\sim$3400--9200 \AA,~which includes the [OIII] doublet\footnote{Other species are considered in Section 3.} up to redshift $\leqslant
0.85$; this
restriction in redshift automatically limits the sample to 28,860 spectra.  The SDSS
pipeline also provides  continuum-subtracted spectra, which in most  cases is quite
acceptable; however, we found  some cases (usually related with the presence of
especially strong and wide Balmer lines) that the estimation of the continuum was
unsatisfactory. To proceed in a consistent way we decided to recalculate the continuum by
fitting a cubic spline to all the 1-d spectra, masking those regions that were centered
around the nominal rest frame wavelengths of strong emission lines. We  inspected several
hundred  spectra  chosen randomly, and checked that this new estimation of the continuum
was reasonably good. Anyway, we have checked that the restrictions found for $\Delta
\alpha/\alpha$ (see next sections) do not change significantly if we use, instead of our
estimation of the continuum, the one provided by SDSS. The wavelength calibration was
checked by measuring the position of the atmospheric OI line. From 1656 sky spectra we
determined a mean position of  $5578.885$ \AA~and a standard deviation of 0.287 \AA~which
agrees quite well with the theoretical value (5578.887 \AA) and with expectations from
the spectral resolution of the spectra. The selection of the final sample was done on the basis of the strength, shape and width
of the [OIII] lines, reliability of the continuum estimation, and possible contribution
due to the relative proximity of the H$\beta$ line. The estimation of the spectral 
position of the [OIII] lines was done following two methods which are described below. 

\subsection{Method 1}
\begin{figure}
\epsscale{0.95}
\includegraphics[scale=0.8]{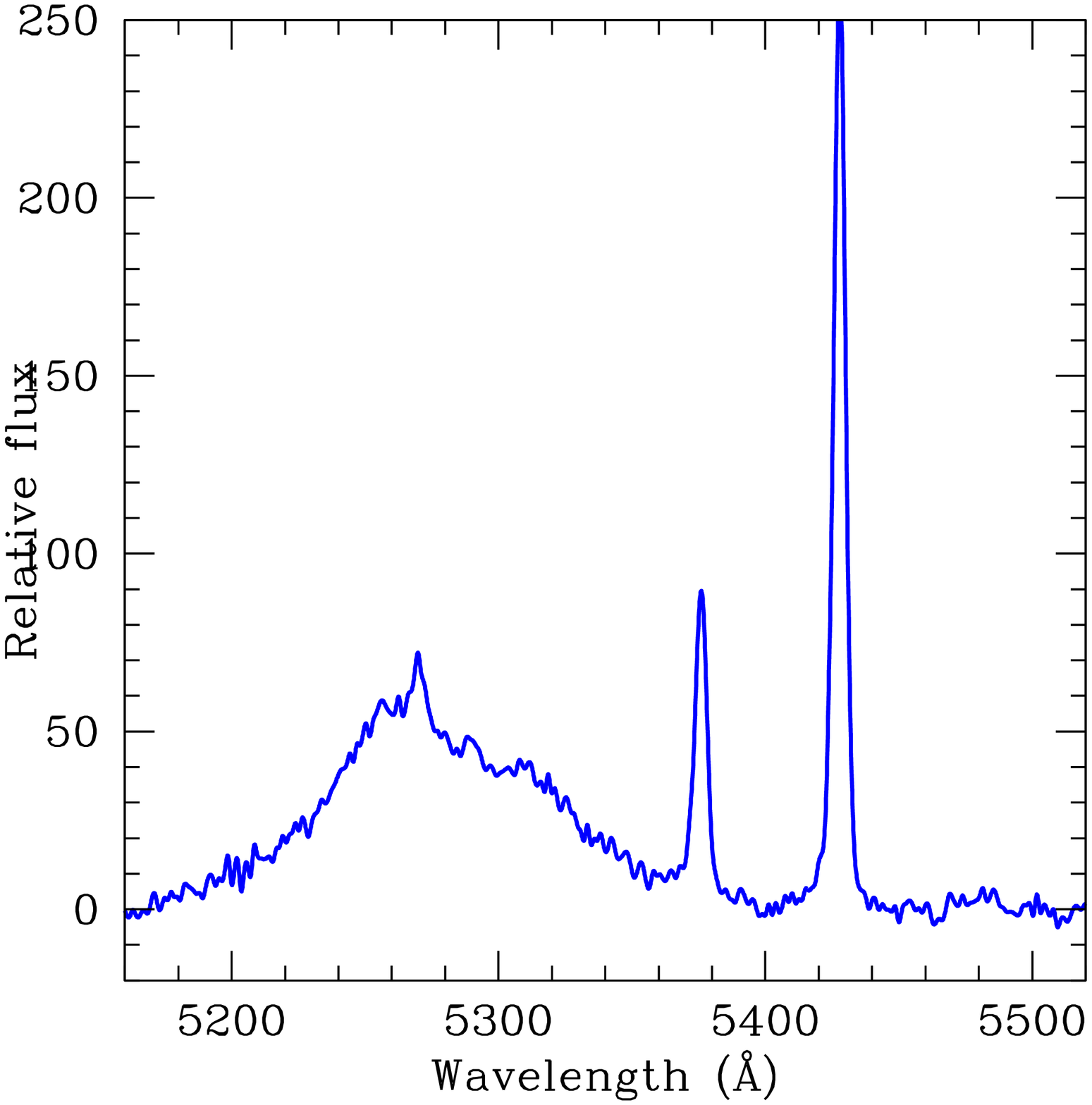}
\caption{Example of one  spectrum that has not been included in the 'clean sample'
on account of its broad  $H\beta$ line.
\label{hbreferee}}
\end{figure}

\begin{figure}
\epsscale{0.95}
\includegraphics[scale=0.8]{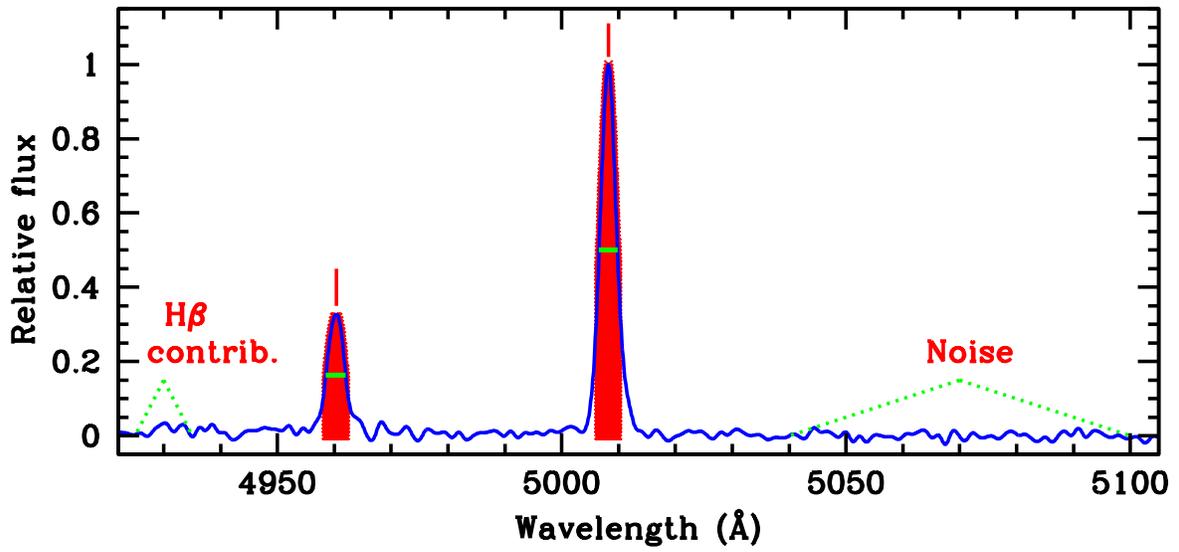}
\caption{Diagram showing the regions used to compute the centroids and  relative fluxes, and the
regions used to assess the $H\beta$ residual contamination and the noise in the continuum.
\label{esquema}}
\end{figure}

\begin{figure}
\epsscale{0.95}
\includegraphics[scale=0.8]{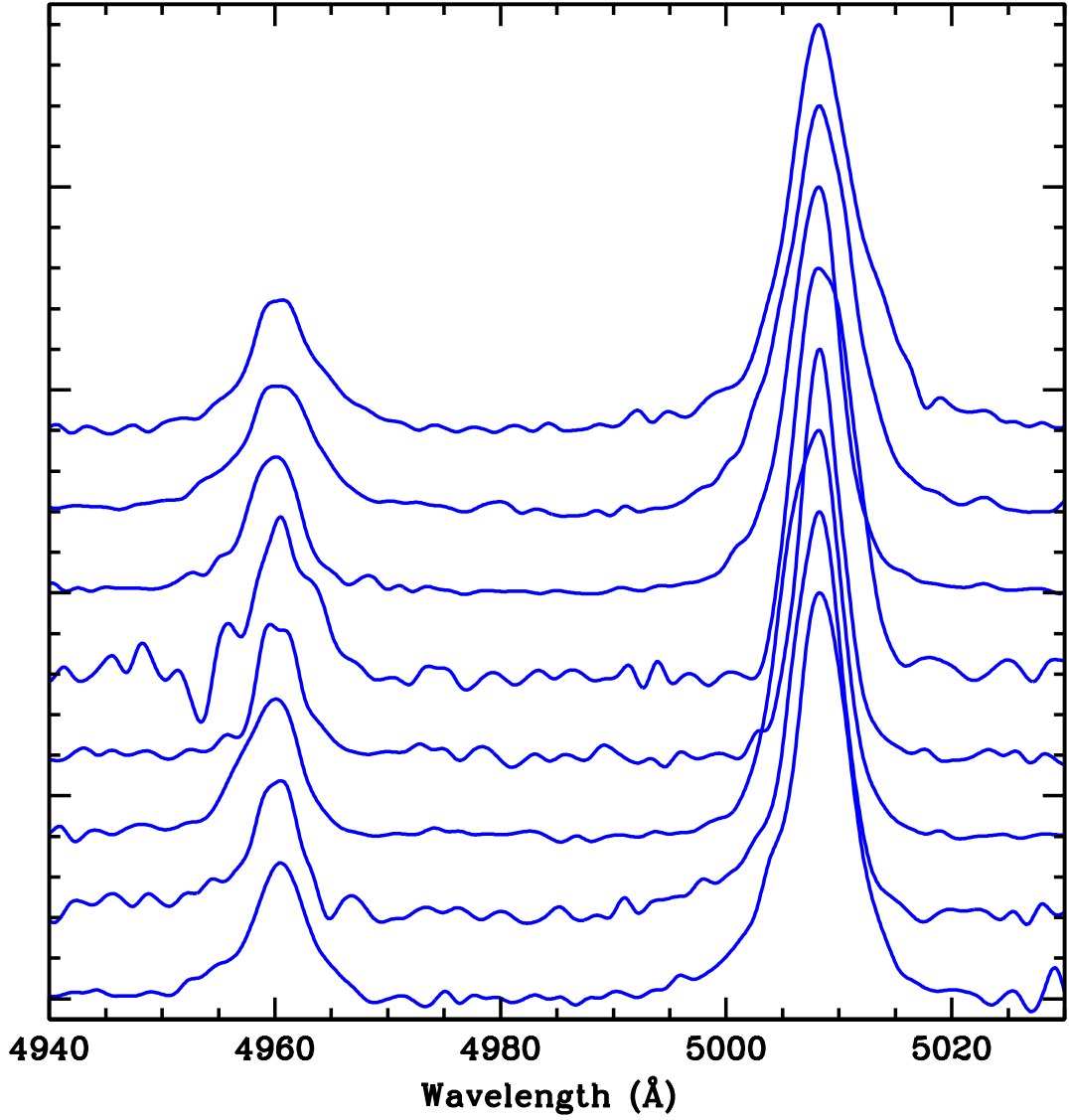}
\caption{The spectral region around the [OIII] doublet for some of the objects included in the clean
sample (see the main text). The spectra have been normalized with respect to the peak of the main
[OIII] line.  \label{examples}}
\end{figure}

To estimate the centroids of the lines, we first determined the FWHM of each
line and averaged the flux of the pixels within the spectral range covered by
that FWHM. We estimate the  relative strength of the [OIII] lines with respect
to the noise in the adjacent continuum. Such noise was estimated by measuring
the continuum rms of each spectrum  in the wavelength range 5040--5100 \AA~(rest
frame). We select those objects that have strong ($SNR>20$ in the peak) [OIII]
lines. These restrictions provide us with a sub sample (which we  refer to
hereafter as the ``raw sample'') of  3739 objects. The contribution of the
H$\beta$ line could in principle produce a blueshift in the estimation of the
centroids of the lines, particularly affecting  the 4959 \AA~line. We checked
that this contribution could be properly quantified by computing the level of
the flux (after subtraction of the continuum) in a spectral region slightly
bluer than the position of the [OIII] ($\lambda\lambda$4959) \AA~line. After
many trials  we chose the   4925--4935 \AA~(rest frame) wavelength range to
estimate possible residuals  of H$\beta$ and remove from the sample those
objects with fluxes in that interval above 0.05 the peak value of the [OIII]
($\lambda\lambda$4959) \AA~line. Fig.\ref{hbreferee} shows an example of a
spectrum that has been rejected on the basis of that criteria. We also eliminated
spectra with very wide or double peaked [OIII] lines.
Figure~\ref{esquema} shows the different spectral regions used to calculate
fluxes and centroids, and to assess potential  $H\beta$ residual contamination
and the continuum level. We do not impose any further  restrictions based on the
shape (the presence of asymmetries) of the [OIII] lines. After all these cuts, 1978 spectra remained to constitute our
``clean 1'' sample. Using this sample we obtain mean $\Delta \alpha/\alpha=(-0.9\pm
2.6)\times 10^{-5}$. The results are very robust against the precise constraints; for instance selecting only those spectra with 
$SNR>30$ in the peak of the [OIII] lines, we obtain mean $\Delta \alpha/\alpha=
(1.9\pm 2.6)\times
10^{-5}$. Figure~\ref{examples} shows examples of the region centered
around the [OIII] doublet of some spectra randomly chosen from this sample. The
figure shows that for a given spectrum both lines are well above the noise
level, and have similar shapes. A few of them show the presence of asymmetries
probably related to the kinematics of the cloud or the presence of multiple
clouds.  The distributions of centroids of each member of the doublet show
similar rms (0.78 \AA); this indicates the absence of significant distortions
in the estimation of the position of the 4959 \AA~line due to the relative
proximity of the H$\beta$ line.  The distribution of the wavelength separation
between the centroids of both lines  has a rms of 0.11 \AA; this indicates that
the main factor broadening the distribution of the centroids of the [OIII] lines 
are absolute spectral shifts which mostly reflect the kinematics ($\le 150$ km
s$^{-1}$) of the narrow line  clouds where the [OIII] originates.

\subsection{Method 2}

\begin{figure}
\epsscale{0.95}
\includegraphics[scale=0.8]{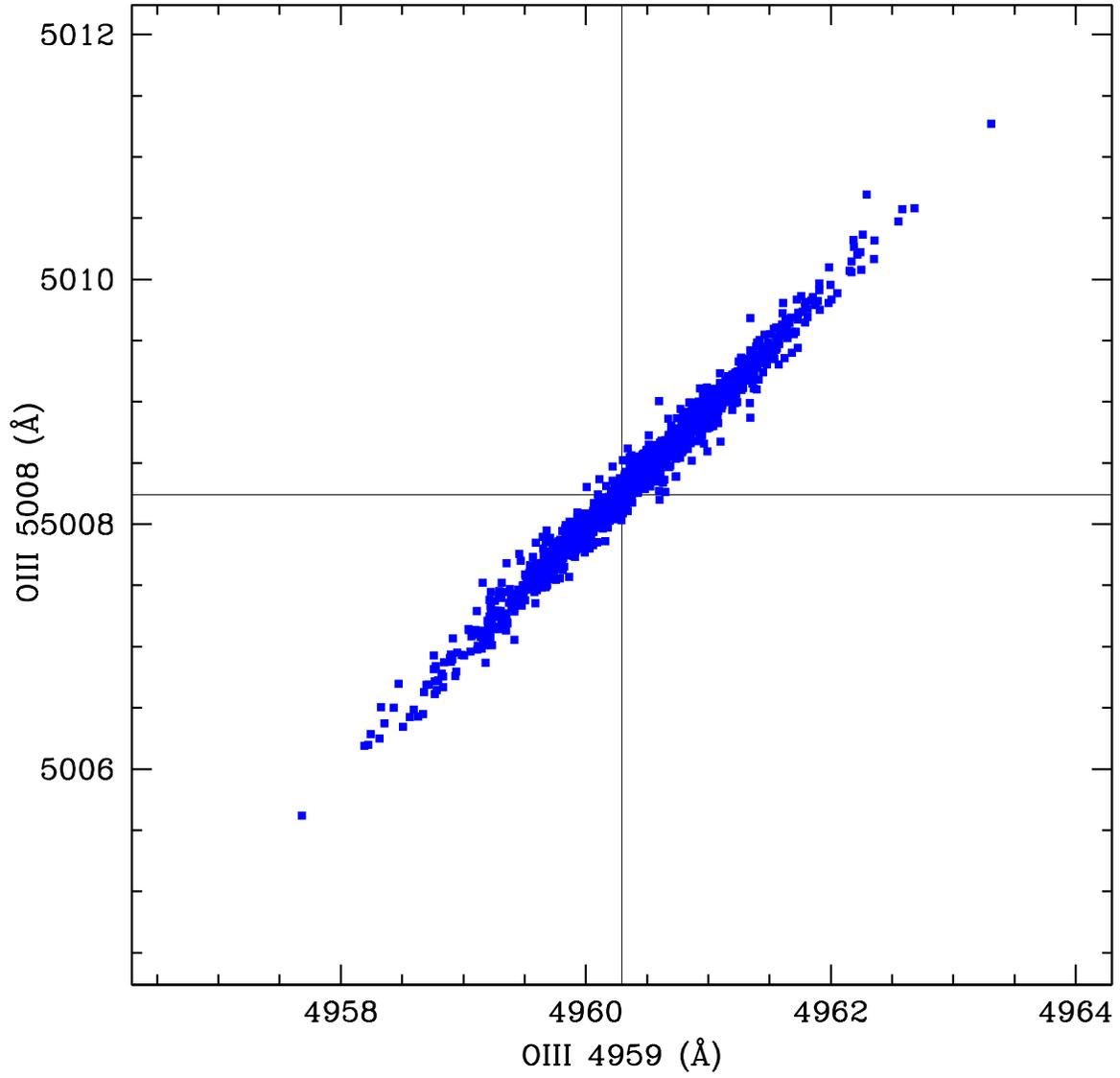}
\caption{Rest frame position of the two lines of the doublet of [OIII] in the clean sample 2 (see
main text). The vertical and
horizontal lines indicate the theoretical local values.\label{o3}}
\end{figure}

The emission lines of OIII (4959 and 5008), and H$\beta$ (narrow and broad
components) were simultaneously modeled by single Gaussian each. The
wavelength position of  each line of the [OIII] doublet was directly estimated
as the central position of the corresponding Gaussian. In principle this seems
a very simple description of the line profiles, and in many cases produces a
poor fit to the data. However, in practice the method takes advantage of the
expected similar shape for both lines of the doublet, and removes most of the
contribution of H$\beta$ in the spectral region of the [OIII] doublet. After
removing the most extreme cases of poor fits, the results are quite consistent
and robust. Selecting those spectra in which each of the [OIII] lines are described by Gaussian with peak
amplitudes $>20$ the level of the continuum noise, and $\sigma$ within the
range 1.4-3.0 \AA~(rest frame of the [OIII] 5008 line) there were 1568 spectra
('clean sample 2') from which we obtained $\Delta \alpha/\alpha=(+2.4\pm
2.5)\times 10^{-5}$.  Figure~\ref{o3} shows the wavelengths determined for the
spectral position of both lines of [OIII] (to compute the rest frame wavelengths
we have used the estimates of redshifts provided by the SDSS database).  The
wavelengths of both lines lie along a line crossing  from the bottom left to
the top right and don't show any systematic effect.

\section{Results and discussion}

The results obtained by both methods are compatible and quite similar. Both estimations are also
compatible with the local value. Most of the spectra included in the clean sample 2 (1223 out of
1568) are also included in the clean sample 1. 

\begin{figure}
\epsscale{0.95}
\includegraphics[scale=0.8]{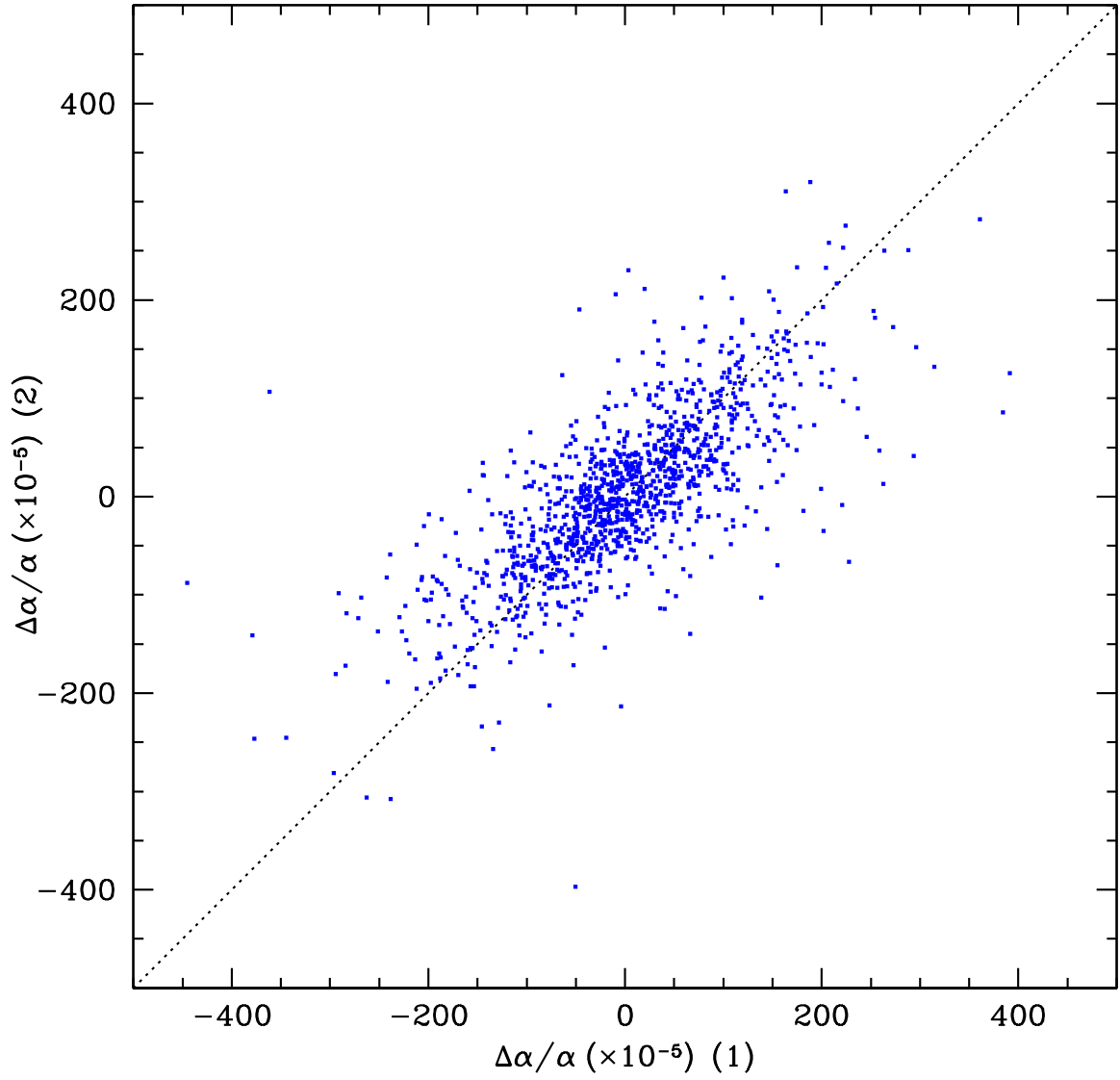}
\caption{Comparison between the estimations of $\Delta \alpha/\alpha$ by each of the methods used
in this paper (see main text for the details). \label{compara_metodos_o3}}
\end{figure}

Fig~\ref{compara_metodos_o3}
shows a comparison between the results of both methods. There is a general good agreement between
them, although the method of fitting Gaussian tends to be a bit more accurate (the rms of the
distributions of $\Delta \alpha /\alpha$ are $9.8\times 10^{-4}$ and $8.0\times 10^{-4}$ for the
clean samples 1 and 2 respectively). Then, here after along the paper, we
will use the results obtained by the second method.  

\begin{figure}
\epsscale{0.95}
\includegraphics[scale=0.8]{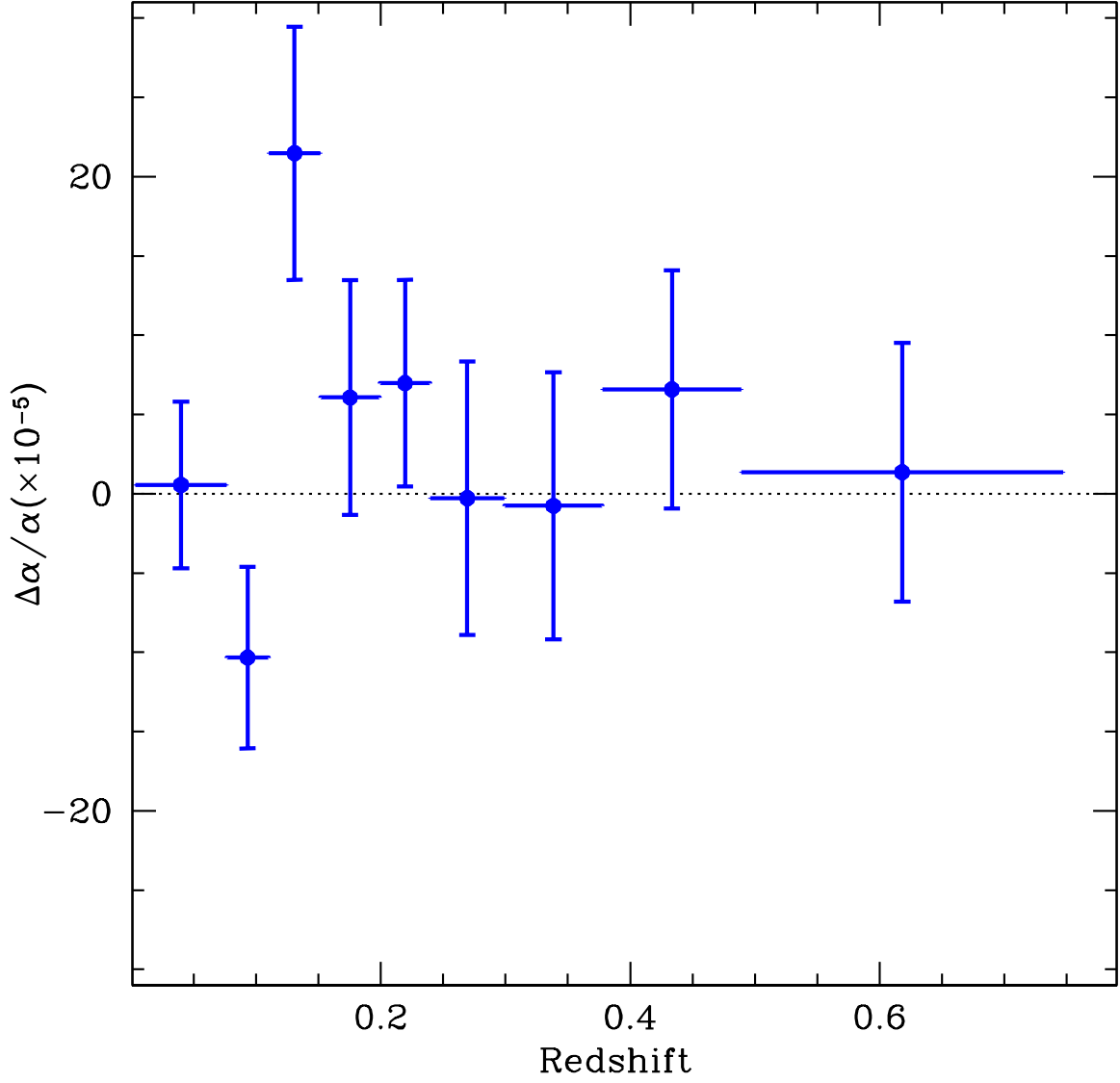}
\caption{$\Delta \alpha/\alpha$ vs. redshift. Each bin contains the contribution of
$\sim 175$ spectra.  
\label{evolz}}
\end{figure}

To analyze the value of $\Delta \alpha/\alpha $ as a function of redshift (or
look-back  time) we compute the mean values in  redshift interval. This is shown in
Fig.~\ref{evolz} and Table~1, which are the main results of this paper. The bins in
redshift have been  built  in order to include approximately the
same number (175) of spectra each. Although none of the bins shows a 
statistical significant departure from zero, the largest deviation is at the
2.7$\sigma$ level in the bin which corresponds to the range in redshift 0.110-0.152.
That bin includes those cases in which one of the [OIII] lines  lies near the
spectral position of the atmospheric OI line. To check for the possible influence of
some residual of the OI line in the estimation of the centroids of [OIII] in that
range of redshift, we built a new sample excluding those spectra having redshifts in
which any  of the [OIII] lines is closer than 15 \AA~to the  spectral position of
the atmospheric OI line. This new constraint removes 59 spectra. The overall results on $\Delta \alpha/\alpha$ do not change
much ($\Delta \alpha/\alpha=(+0.6\pm 2.4)\times 10^{-5}$) using this new clean
sample, but the  significance of the departure from zero in the corresponding bin 
in redshift (0.107-0.161) is largely reduced ($\Delta \alpha/\alpha=(+7.0\pm
5.8)\times 10^{-5}$). 

The range in redshift spanned by our sample corresponds to a
maximum look-back time of 6.6 Gyr, and a mean of 3.0 Gyr. Following  Bahcall et
al., the mean rate of possible changes is $1/ <t>\Delta \alpha/\alpha=(+0.7\pm
0.7)\times 10^{-14}$ yr$^{-1}$. A linear fit of  $\Delta \alpha/\alpha$ with
respect to look-back time gives a slope of $(0.2\pm 1.6)\times 10^{-14}$
yr$^{-1}$.  All these numbers as a whole do not show any significant evidence of
change in $\Delta \alpha/\alpha$ with redshift. 

\subsection{Line ratios}

\begin{figure}
\epsscale{0.95}
\includegraphics[scale=0.8]{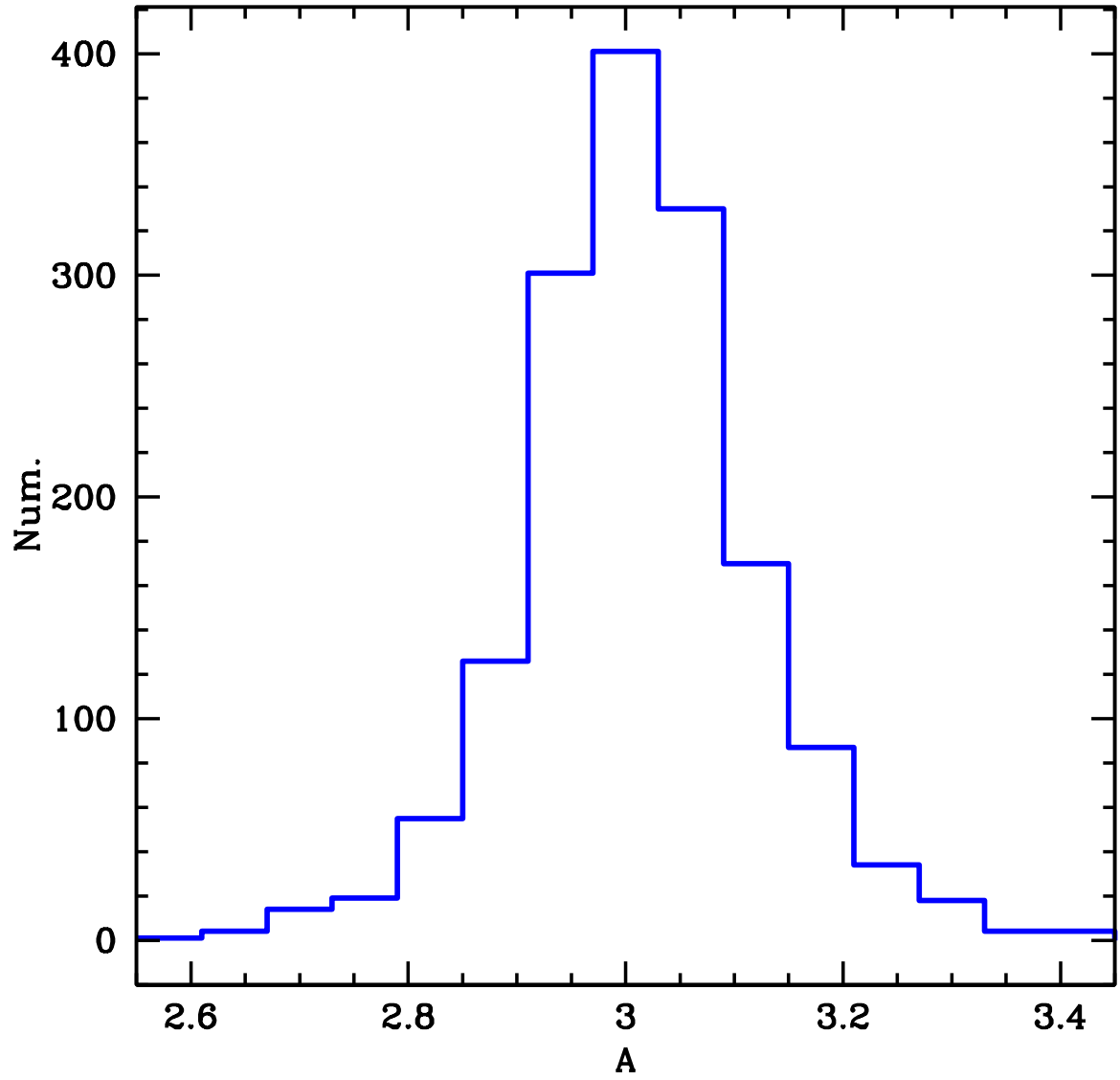}
\caption{Histogram of the line ratio A of the [OIII] lines.
\label{coef}}
\end{figure}

As a by-product of the analysis we have estimated the value of the line ratio of [OIII]:

$$A\propto \frac{A(5008)\Delta E(5008)}{A(4959)\Delta E(4959)}$$ 

\noindent where $A$ and $\Delta E$ are the transition rates and the energy
differences, respectively. The determination of this ratio
is interesting as an observational test of calculations from atomic theory where a
$\sim 3$ \% level of discrepancy exists among the published results \citep{gru05}.  
We performed numerous tests checking the dependence of $A$ on possible
residual contamination from $H\beta$ (see previous section), SNR and  [OIII] width.
Our best estimation is obtained with the same constraints as those used for the
estimation of $\Delta \alpha/\alpha$ using method 2. Fig.~\ref{coef} shows the
histogram of $A$ values for the 1568 spectra included in 'clean sample 2'; the
resulting distribution is symmetric, centered at $\sim 3$ and  has a
standard deviation of 0.10. From that distribution was obtained our best estimation of
$A=3.012\pm 0.003\pm 0.010$, where the error bars  correspond to
the 1$\sigma$ statistical and  systematic errors respectively. The systematic errors
correspond to uncertainties due to the criteria for selecting the samples. The value
obtained agrees with the estimate by Bahcall et al. ($2.99\pm 0.02$), with the
measurements  $2.993\pm 0.014$ by \citet{dim07} using 62 AGN, and it is
compatible with theoretical estimates (2.98) by \citet{sto00}.

\subsection{Other doublets}

As  pointed out by Bahcall et al.\ and developed and used by Grupe et al., it
is possible to constrain the value of $\alpha$ at cosmological look-back times
using doublets of other emission lines that are common in the spectra of QSOs.
The  most promising are [NeV] ($\lambda \lambda 3346, 3426$), [Ne III]
($\lambda \lambda 3869, 3968$), [OI] ($\lambda \lambda 6300, 6365$), and [SII]
($\lambda\lambda 6717, 6731$). The main advantages of the use of these lines is
the possibility of extending the study to higher redshifts and of checking  the
consistency of the results using the [OIII] lines. All of them lie in the
optical range  and,  in the wavelength range of SDSS, can be useful for
exploring the ranges in redshifts up to 1.69, 1.31, 0.44 and 0.37 respectively.
However, these lines are usually weaker than [OIII] in QSO spectra and some of
them also have  drawbacks that limit their suitability for these studies. For
instance, the lines of the [SII] doublet are relatively close so that further
constraints on the width of such lines need to be applied. This pair is also
relatively close to the usually very strong and wide H$\alpha$ line in QSOs. On
the other hand, the pair of lines for  [NeV] and [NeIII] are separated by
$\sim$100 \AA~and then spectral calibration uncertainties might be a more
serious concern. 

\begin{figure}
\epsscale{0.95}
\includegraphics[scale=0.8]{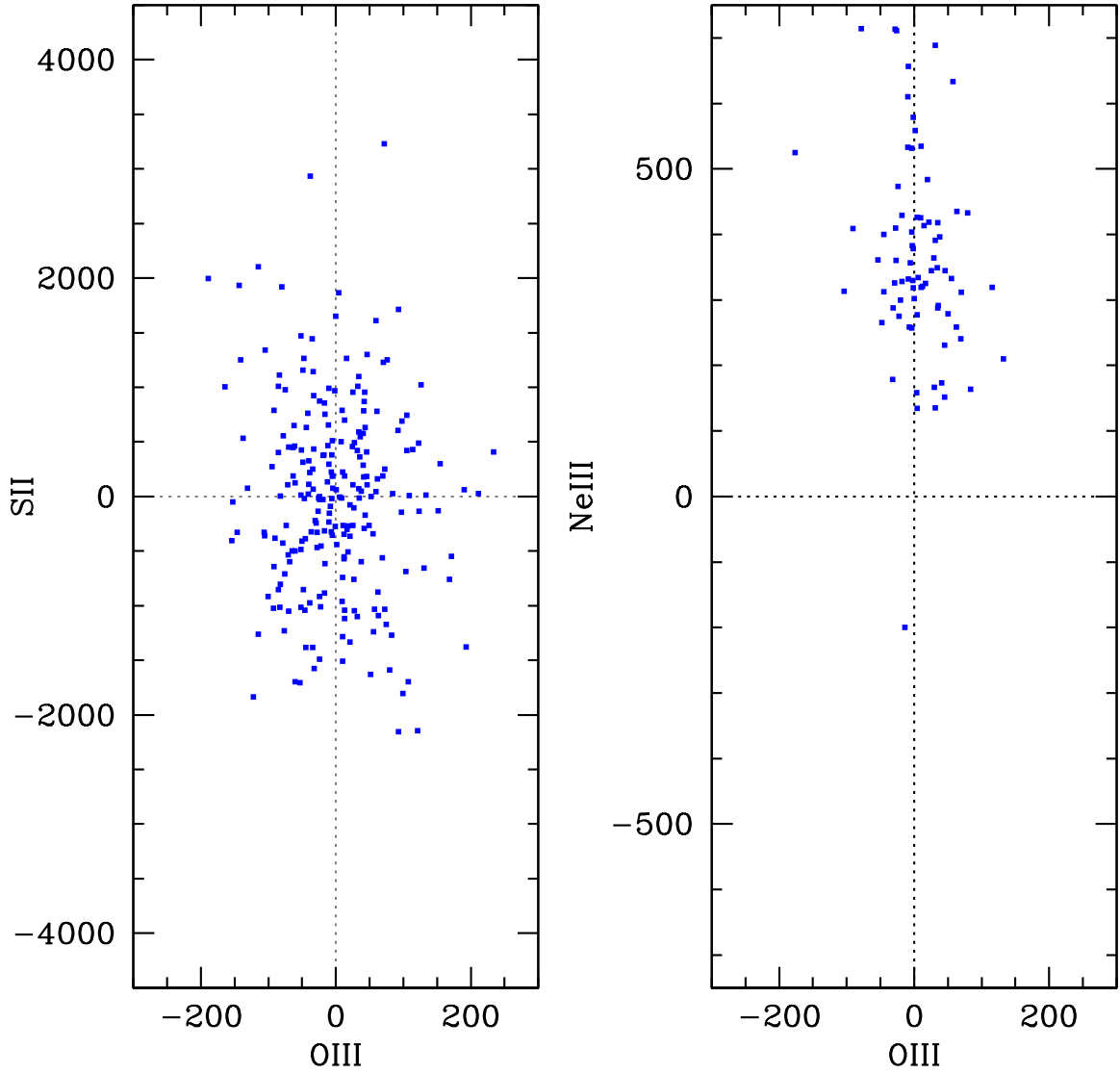}
\caption{Comparison between the estimations of  $\Delta \alpha/\alpha$  obtained
from [OIII] and those derived from [SII] ($left$) and [NeIII] ($right$) respectively.
\label{compara_especies}}
\end{figure}

We follow a similar procedure to the one used for [OIII], adapting the
constraints on the SNR, possible contributions from neighboring lines, etc., to
each specific doublet.  As expected, because of the relatively low amplitude of
such lines with respect to [OIII], the clean samples of each of these pairs are
much smaller, and  the sensitivity achieved  is poorer than  with [OIII]. We 
decided not to use the  [NeV] and [OI] doublets for this analysis  because the
number of spectra with lines strong enough for a meaningful analysis is very
small. In order to have a representative sample of [NeIII] and [SII] we relax
our selection criteria to include in the samples all the spectra having  SNR$
>10$ in the peak of each line of the doublet. Table~2  presents the results for
[NeIII] and [SII]. The values found from [SII] are compatible with those from
[OIII].  Although the clean sample used in the case of the [SII] contains a
relatively high number of spectra (466), the sensitivity achieved is
comparatively much lower than the one obtained from [OIII] and [NeIII].  After
doing exhaustive tests we found that the relative separation with respect to
the width of the lines is the most important factor which limits the use of
this pair.  Figure~\ref{compara_especies} $(left)$ shows a comparison between the $\Delta \alpha/\alpha$
values obtained in common spectra between [SII] and [OIII] samples. The figure shows the compatibility of
both estimations  and the comparatively much large dispersion obtained from the estimations based on [SII].

The values of $\Delta \alpha/\alpha$  obtained from [NeIII] are
significantly different from zero. We have checked that this departure from
zero   keeps along all the range in redshift analyzed using that
doublet.  The wide range in redshift covered by this sample and the absence of
the effect for other lines (for instance, the lines of the [OIII] doublet or the
atmospheric OI line) precludes  possible systematic errors in wavelength
calibration. Figure~\ref{compara_especies} ($right$) shows a comparison between
the estimations of $\Delta \alpha/\alpha$ from the [OIII] and [NeIII]
respectively using only common spectra  to both samples. Clearly, the
results from [NeIII] are not compatible with those from [OIII]. For the
distribution of the centroids of each member of the [NeIII] doublet, we have
estimated  mean values of 3869.61 and 3969.05 \AA~and rms of 0.51 and 0.58 \AA~
respectively. The comparison  with the theoretical local values (3869.86 and
3968.59 \AA~respectively) shows discrepancies of 0.25 and 0.46 \AA~to the
blue and to the red respectively. If we  restrict the sample according to 
stronger constraints  on the SNR of both lines, these shifts tend to decrease
only  for the line at 3870 \AA. Although this could indicate inaccuracy in the
tabulated wavelengths of both transitions, we think the main reason is that the
estimated wavelengths for the line [NeIII] (3970 \AA) are affected by the
proximity  of several
lines like
H$\epsilon$ and the absorption line of CaII-H. Nevertheless, if this is the case it
would remain unclear why the rms of the the distributions of both members of the
doublet are so similar, and why the shift in the 3970 \AA~line does not decrease
when only spectra with relatively high SNR are considered.  Curiously enough, the high
value found for  $\Delta \alpha/\alpha$ using [NeIII]  is similar to the mean
value found by Grupe et al. using this pair.

\subsection{Comparison with previous studies}

\begin{figure}
\epsscale{0.95}
\includegraphics[scale=0.8]{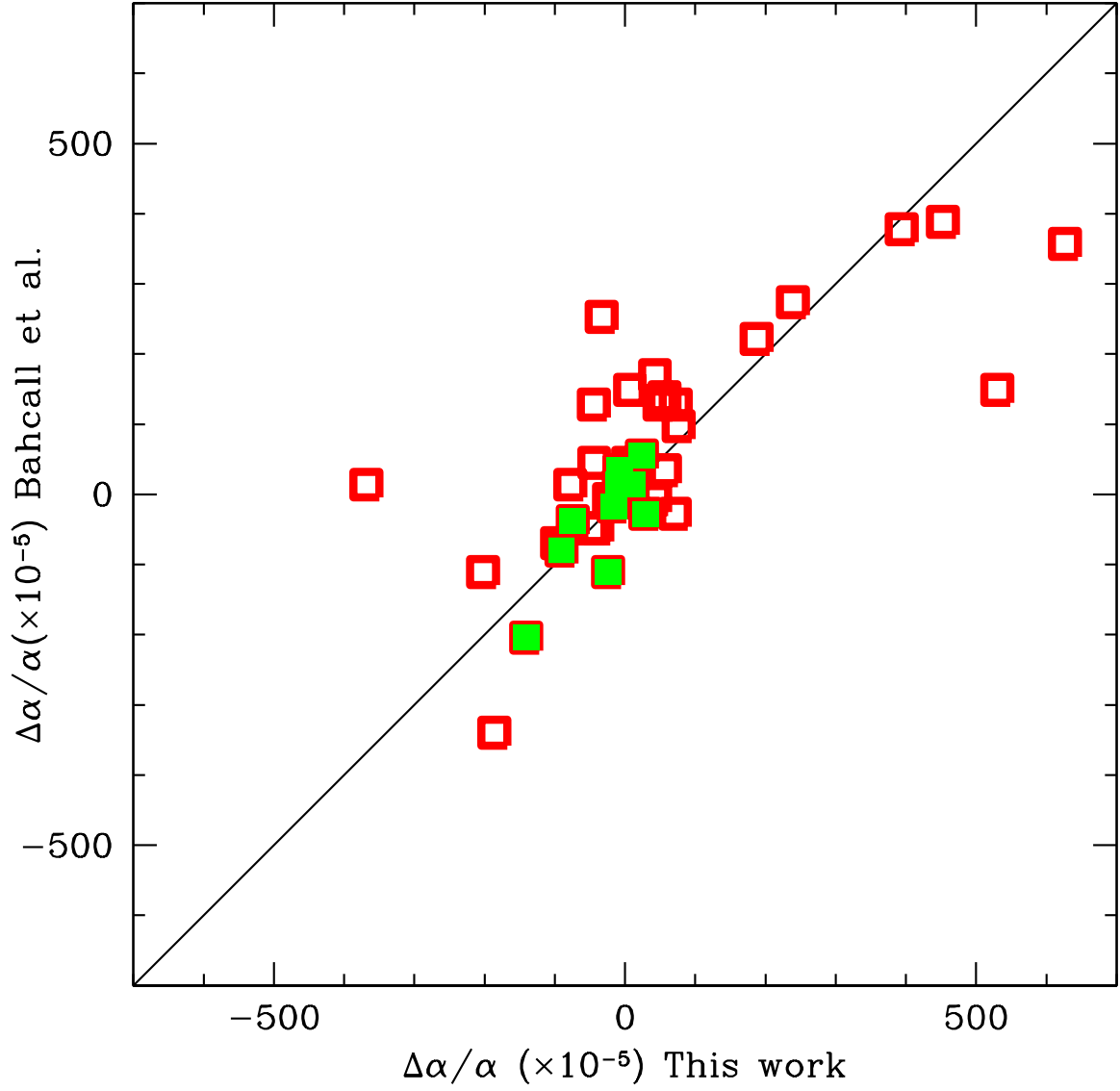}
\caption{Comparison between the values of $\Delta \alpha/\alpha$ estimated by Bahcall et al. and
in this study. Results from spectra included in the raw (clean) sample are plotted as
empty (solid)
points.
\label{bahcall}}
\end{figure}

In Section~1 we have mentioned the different astronomical methods used to constrain the
variation in $\alpha$ with particular emphasis on those based on  fine structure
splitting. Here, we make comparisons only with previous observations based on the
relative splitting of pairs of emission lines in QSOs or AGN in general. The analysis by
Bahcall et al.\ is based on the Early Data Release of SDSS, which comprises a sub sample
of the sample analyzed in this paper, so it is worth  making a comparison of both
studies. There are 38(10) shared objects between the sample of these authors and our
raw(clean) sample. Figure~\ref{bahcall} shows a comparison between our estimates of
$\Delta \alpha/\alpha$ and those by Bahcall et al. The comparison shows a very good
agreement between both studies. 

The distribution of $\Delta \alpha/\alpha$ obtained from objects  in common between our
raw sample and the Bahcall et al.\ sample is similar, while the rms of $\Delta
\alpha/\alpha$ from our clean sample is $0.5\times 10^{-3}$, slightly better than the sample of
Bahcall et al.\ ($0.7\times 10^{-3}$). So, although the method and criteria for selecting
the samples are quite different from each other, both studies agree quite well. The much higher number of spectra used in this
study  (1568 in our clean sample) with respect to the 42 objects in the Bahcall et al.\
sample would allow us in principle to reduce the uncertainty by a factor
$\sqrt{1568/42}=6.1$. The result presented by Bahcall et al.\footnote{In the appendix of the Bahcall et al. paper  the SDSS Data Release One
was analyzed and a value $\Delta \alpha/\alpha=(1.2\pm 0.7)\times 10^{-4}$ found.} is
$(0.7\pm 1.4)\times 10^{-4}$ , and
therefore our result ($\Delta \alpha/\alpha =(+2.4\pm
2.5)\times 10^{-5}$) improves the sensitivity by a factor $\sim$5,
which agrees with statistical expectations.

\citet{gru05} applied the same method to a sample of 14 Sey 1.5 galaxies up to
redshift 0.281. These authors include in their analysis the [NeIII], [NeV], [OIII], [OI],
and [SII] doublets.  They found $\Delta \alpha/\alpha=(1.5\pm 0.7)\times 10^{-3}$, so the
result presented  in this study improves that sensitivity by a factor $\sim 30$, and the
range in redshift covered by a factor $\sim 3$. The method  used in this paper  to constrain $\alpha$ in the past is about one order of
magnitude less sensitive than methods based on the analysis of absorption systems along
the line of sight of QSOs. However, as has been discussed above, the method is more
transparent, does not require assumptions on the spatial distribution and kinematics of
the gas involved, and in general is less subject to systematic errors. 

\section{Conclusions}

The main findings of this work are:

\begin{itemize}

\item We have analyzed the sample of 77092 objects classified as QSOs in the SDSS DR6, and from
these we have selected a sample of 1568 objects with strong [OIII] emission  lines up to redshift
0.80.

\item From the relative spectral position in such spectra of the nebular emission  of [OIII] $\lambda\lambda
5007$ and 4959 we have estimated a mean value for the fine structure constant 
of $\Delta \alpha /\alpha =(+2.4\pm 2.5)\times 10^{-5}$ through the last 6.6 Gyr. 

\item Our analysis improves by a factor $\sim 5-6$ the existing uncertainty in the estimates  in
the possible change of $\alpha$ using the same method.

\item The  maximum rate of change in $\Delta \alpha/\alpha$ allowed by our analysis is  
$(+0.7\pm 0.7)\times 10^{-14}$ yr$^{-1}$ in the last 6.6 Gyr.

\end{itemize}

\acknowledgments MLC  was supported  by the {\it Ram\'on y Cajal} Programme of the
Spanish science ministry. We have used the following online databases: Sloan Digital Sky
Survey (http://www.sdss.org/), the Atomic 
Line List (http://www.pa.uky.edu/~peter/atomic/), and NIST Atomic Spectra Database
(http://physics.nist.gov/PhysRefData/ASD).

\clearpage

\clearpage

\clearpage

\begin{figure}
\epsscale{0.95}
\includegraphics[scale=0.8]{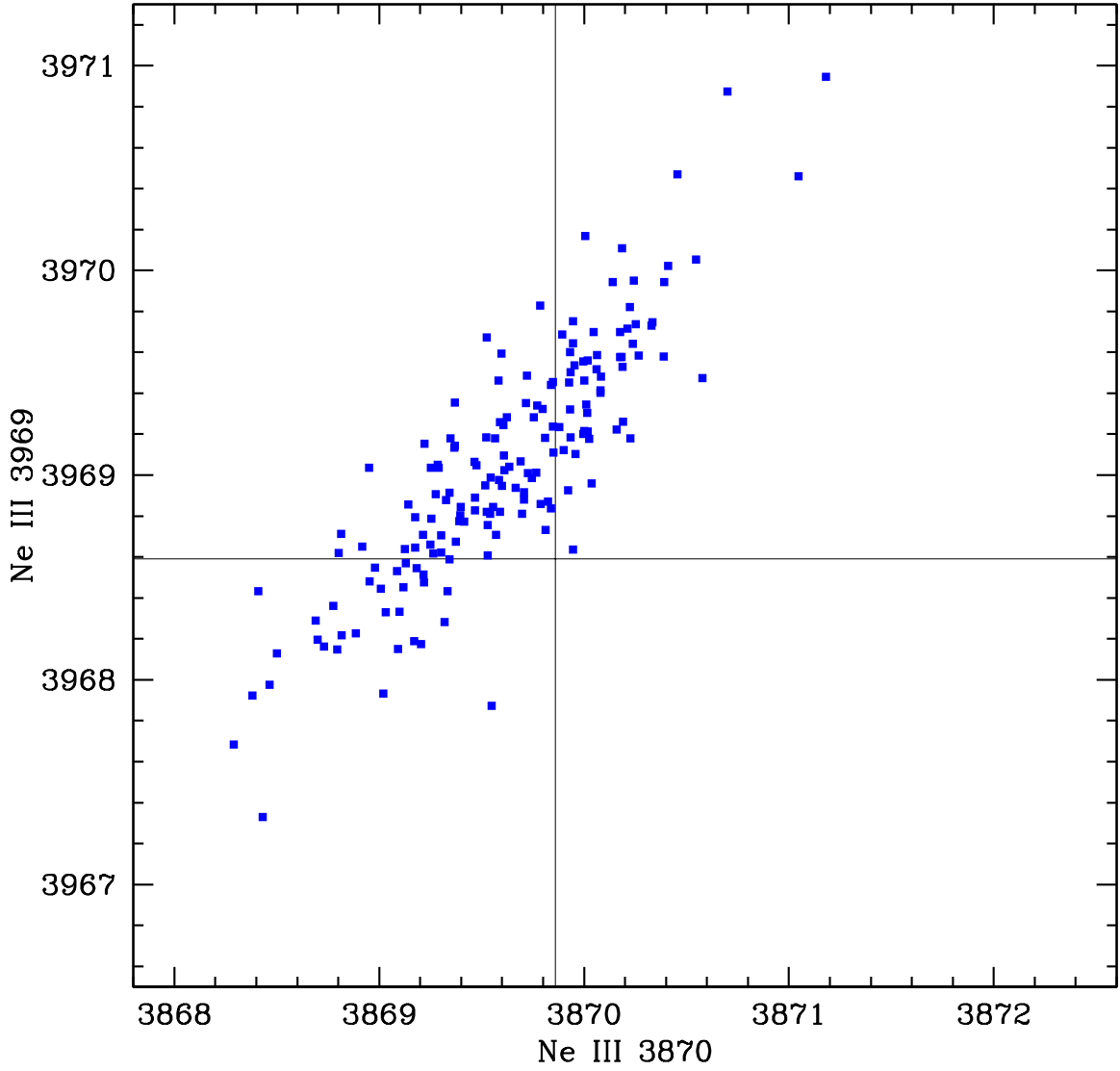}
\caption{Rest frame position of the two lines of the doublet of [NeIII]. The vertical and
horizontal lines indicate the theoretical local values.\label{parejas_ne3}}
\end{figure}

\clearpage

\begin{table}
\begin{center}
\caption{$\Delta \alpha/\alpha$ at different redshifts}
\begin{tabular}{lrr}
Redshift & $\Delta \alpha/\alpha$ ($10^{-5})$ &  1-$\sigma$ errors  \\
\tableline
 0.003  -      0.076    &	   +0.6        &  5.2 \\
 0.076  -      0.110    &	 -10.3        &  5.7 \\
 0.110  -      0.152    &	  +21.5        &  8.0 \\
 0.152  -      0.199    &	   +6.1        &  7.4 \\
 0.199  -      0.240    &	   +7.0        &  6.5 \\
 0.240  -      0.300    &	  -0.3        &  8.6 \\
 0.300  -      0.378    &	  -0.8        &  8.4 \\
 0.378  -      0.490    &	   +6.6        &  7.5 \\
 0.490  -      0.747    &	   +1.4        &  8.2 \\ 				      
\tableline					 
\end{tabular}
\end{center}
\end{table}

\begin{table}
\begin{center}
\caption{Results for the different atomic species}
\begin{tabular}{lrrr}
Doublet & Spectra & $\Delta \alpha/\alpha$ ($10^{-5})$ &  1-$\sigma$ errors \\
\tableline
OIII  & 1568  &    +2.4  &   2.5 \\
NeIII &  168  &  +358.5  &  11.2 \\
SII   &  481  &  -18.8  &  41.8 \\
\tableline
\end{tabular}
\end{center}
\end{table}
\end{document}